\documentclass[reprint,amsmath,amssymb,aps,superscriptaddress]{revtex4-1}

\usepackage{amsmath}
\usepackage{graphicx}
\usepackage{dcolumn}
\usepackage{bm}
\usepackage{booktabs}
\usepackage{array}
\usepackage[capitalise]{cleveref}
\usepackage{multirow}

\usepackage[detect-weight=true, detect-family=true]{siunitx}

\usepackage{tikz}
\tikzstyle{box}=[rectangle,rounded corners, draw=black, very thick,text width=2.3cm,minimum height=3em,text centered]
\usepackage{pgfplots}
\usepgfplotslibrary{groupplots}
\usepgfplotslibrary{colorbrewer}
\usepackage{pgfplotstable}
\pgfplotsset{compat=1.13}
\usepackage{pgffor}

\usepackage{subcaption}

\usepackage{comment}

\definecolor{tud1d}{RGB}{36,53,114}

\begin{document}
\preprint{APS/123-QED}

\title{Uncertainty Modeling and Analysis of the European X-ray Free
Electron Laser Cavities Manufacturing Process}

\author{J. Corno}
\affiliation{Technische Universit{\"a}t Darmstadt, Darmstadt, Germany}

\author{N. Georg}
\email{n.georg@tu-braunschweig.de}
\affiliation{Technische Universit{\"a}t Darmstadt, Darmstadt, Germany}
\affiliation{Technische Universit{\"a}t Braunschweig, Braunschweig, Germany}

\author{S. Gorgi Zadeh}
\author{J. Heller}
\affiliation{Universit{\"a}t Rostock, Rostock, Germany}

\author{V.  Gubarev}
\affiliation{DESY, Hamburg, Germany}

\author{T. Roggen}
\affiliation{CERN, Geneva, Switzerland}

\author{U. R{\"o}mer}
\affiliation{Technische Universit{\"a}t Braunschweig, Braunschweig, Germany}

\author{C. Schmidt}
\affiliation{Universit{\"a}t Rostock, Rostock, Germany}

\author{S. Sch{\"o}ps}
\affiliation{Technische Universit{\"a}t Darmstadt, Darmstadt, Germany}

\author{J. Schultz}
\affiliation{Technische Universit{\"a}t Braunschweig, Braunschweig, Germany}

\author{A. Sulimov}
\affiliation{DESY, Hamburg, Germany}

\author{U. van Rienen}
\affiliation{Universit{\"a}t Rostock, Rostock, Germany}
\affiliation{ Department Life, Light \& Matter (LL\&M), Universit{\"at} Rostock, Rostock, Germany}

\begin{abstract}
This paper reports on comprehensive efforts on uncertainty quantification and global sensitivity analysis for accelerator cavity design. As a case study object the TESLA shaped superconducting cavities, as produced for the European X-ray Free Electron Laser (EXFEL), are selected. The choice for these cavities is explained by the available measurement data that can be leveraged to substantiate the simulation model. Each step of the manufacturing chain is documented together with the involved uncertainties. Several of these steps are mimicked on the simulation side, e.g. by introducing a random eigenvalue problem. The uncertainties are then quantified numerically and in particular the sensitivities give valuable insight into the system behavior. We also compare these findings to purely statistical studies carried out for the manufactured cavities. More advanced, adaptive, surrogate modeling techniques are adopted, which are crucial to incorporate a large number of uncertain parameters. The main contribution is the detailed comparison and fusion of measurement results for the EXFEL cavities on the one hand and simulation based uncertainty studies on the other hand. After introducing the quantities of physical interest for accelerator cavities and the Maxwell eigenvalue problem, the details on the manufacturing of the EXFEL cavities and measurements are reported. This is followed by uncertainty modeling with quantification studies.
\end{abstract}
\maketitle

\section{INTRODUCTION}

Accelerator devices require advanced, simulation based, design approaches due to demanding performance requirements and a considerable level of technical complexity. This is particularly true for superconducting accelerator cavities, which are a key element of many modern accelerator facilities. A typical design process involves 2D as well as 3D numerical solutions of the Maxwell eigenvalue problem, followed by optimization and uncertainty analysis and quantification studies. The latter have been conducted within the accelerator community from the 1970s, \cite{Halbach_1976aa, Weiland_1977, vanRienen_1987}. However, these studies have been mainly based on (local) sensitivity analysis which should be applied with care to quantify uncertainties in the cavities' geometry. Indeed, the eigenmodes and other measures of interest depend strongly on the shape of the cavity and local measures may not yield reliable results. 

The topic of simulation based uncertainty quantification has seen tremendous developments in recent years, also in computational electromagnetics, see e.g. \cite{Clenet_2013aa}. Nowadays, significant computational resources are available and uncertainty studies, taking into account systematically large parameter variations at all steps of the design process, come into reach. In particular, the concept of global sensitivity analysis \cite{saltelli2008global} has received much attention. The variance-based approach to global sensitivity analysis, measures the contribution of each parameter (or parameter combination) to the variance of a system output quantity. These Sobol sensitivity indices permit not only to analyze the importance of model input parameters, which in turn is useful in guiding modeling efforts, but also to identify important combined high-order parameter variations.  Despite their clear interpretation, Sobol coefficients are not readily extendable to settings with correlated inputs, which are useful in our case to model several parts of the manufacturing and assembly process. Such effects in turn can be quantified by Borgonovo indices \cite{Borgonovo_2007} which represent another approach to global sensitivity analysis, which has not received much attention in physics and engineering applications so far. We will apply both methods in the present study, however, their derivations will only be recalled in appendix~\ref{app:Sobol}. Although the concept of global sensitivity analysis is quite well-established, the efficient computation of sensitivity indices for cavity applications is a difficult task, mainly due to the complexity of the underlying eigenvalue problem \cite{Georg_2019aa}.  This is addressed here by introducing surrogate models which emulate the relation between eigenmodes, or other quantities of interest on the model parameter. 

In this respect, this paper reports on comprehensive efforts on uncertainty quantification and global sensitivity analysis for the simulation of TESLA shaped, superconducting, cavities. Such cavities have been produced in considerable quantity for the European X-ray Free Electron Laser (EXFEL) and measurement data is available to substantiate the approach. Each step of the manufacturing chain is documented together with the involved uncertainties. Some of these steps are mimicked on the simulation side, e.g. by introducing a random eigenvalue problem. The uncertainties are then quantified numerically and in particular the sensitivities give valuable insight into the systems behavior. We also compare these findings to purely statistical studies carried out during the manufacturing. However, the simulation of all manufacturing steps would require the solution of several random inverse problems and is postponed to future work. 

Uncertainty studies in an accelerator physics context have been reported before, see \cite{Xiao_2007aa,Schmidt_2014aa,Corno_2015ac}. In this work, we use more advanced, adaptive, surrogate modeling techniques, which are crucial to incorporate a large number of uncertain parameters. The main contribution is the detailed comparison and fusion of real manufacturing data on the one hand and simulation based uncertainty studies on the other hand. The paper also clearly points out important directions of future research, which would allow to further combine measurements and simulation.

The structure of the paper is given as follows: Section \ref{sec:quantities_of_interest} introduces quantities of physical interest for accelerator cavities and the Maxwell eigenvalue problem. In Section \ref{sec:manufacturing}, details on the manufacturing of the EXFEL cavities and measurements are reported. This is followed by uncertainty modeling and quantification studies in Section \ref{sec:uncertainty} and concluding remarks.

\section{Cavities and Maxwell Eigenvalue Problem}
\label{sec:quantities_of_interest}

\begin{figure}[hb]
\includegraphics[width=.7\columnwidth]{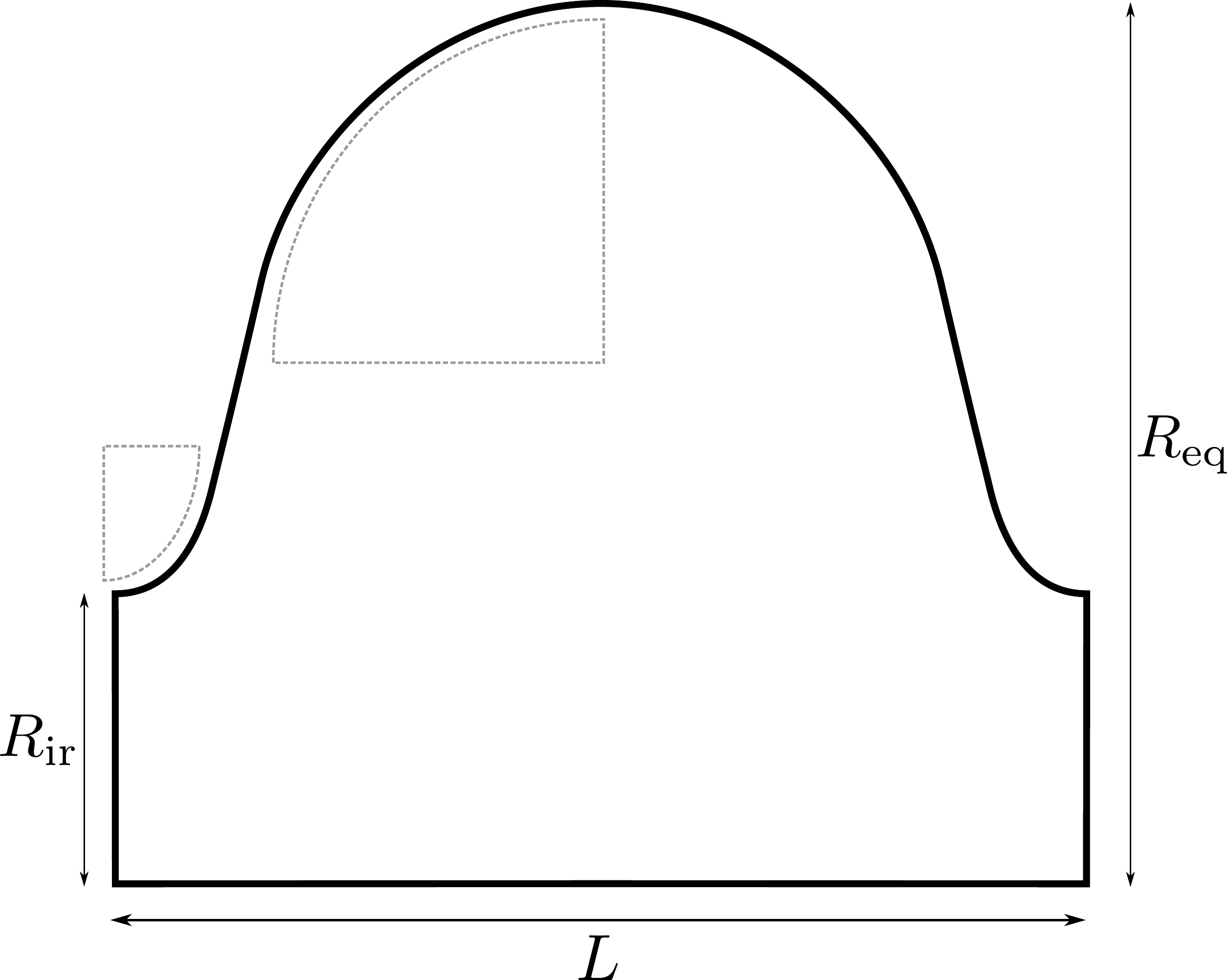}
\caption{Geometrical shape of an elliptical cell.}
\label{fig:cavity_params}
\end{figure}

Accelerating cavities are devices used to accelerate particles to higher energies. Elliptical cavities are the accepted geometrical shape for particle velocities close to the speed of light ($\beta\approx 1$, where $\beta$ is the ratio of particle velocity to the speed of light in vacuum). The shape of an elliptical cell is defined by two elliptical arcs connected by a tangent straight line as shown in FIG.~\ref{fig:cavity_params}. The fundamental mode of the cavity is the TM$_{010}$ mode that is typically used as the operating mode of the cavity. 

\begin{figure}[t!]
\includegraphics[width=\columnwidth]{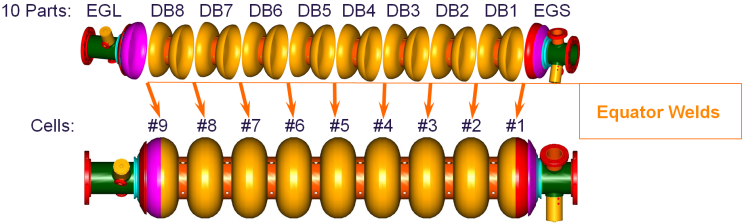}
\caption{The European XFEL cavity and its parts: 8 dumb-bells (DB) and end-groups (EGS, EGL).}
\label{fig:xfel_cavity}
\end{figure}

In order to enhance the accelerating efficiency, multi-cell cavities are created by connecting several cells together via their irises (see FIG.~\ref{fig:xfel_cavity} for the EXFEL cavity described in more detail in Section \ref{sec:manufacturing}). In a multi-cell cavity of $N_\text{c}$ cells, there are $N_\text{c}$ modes of each type due to multi-cell coupling. In this work, we are interested in the TESLA cavity shape \cite{Aune2000}, which is composed of $N_\text{c}=9$ cells. 

Let $\Omega(\boldsymbol{Y}) \subset \mathbb{R}^3$ refer to the inner domain of the multi-cell cavity with boundary $\partial\Omega(\boldsymbol{Y})$, where $\boldsymbol{Y}$ denotes a vector of shape parameters to be specified. The fields in the source-free, 
time-harmonic case, are given by Maxwell's equations:
\begin{equation}\label{eq:Maxwell-time-harmonic}
\begin{aligned}
\nabla\times{\mathbf{E}}       &= -j\omega\mu_0\mathbf{H} 
&\quad
\nabla\times{\mathbf{H}}       &= j\omega\epsilon_0\mathbf{E}\\
\nabla\cdot{\epsilon_0\mathbf{E}} &= 0 
&
\nabla\cdot{\mu_0\mathbf{H}}  &= 0,
\end{aligned}
\end{equation}
where $\mathbf{E}$ and $\mathbf{H}$ denote the electric and magnetic field strength, $\varepsilon_0$ and $\mu_0$ are the permittivity and permeability of  vacuum, respectively. The walls are modeled as perfect electric conductor, i.e.,
\begin{equation}\label{eq:pecBC}
\begin{aligned}
\mathbf{E}\times\mathbf{n} &= 0   &\quad
\mathbf{H}\cdot\mathbf{n}  &= 0.
\end{aligned}
\end{equation}
One derives the Maxwell eigenvalue problem from \eqref{eq:Maxwell-time-harmonic}
by eliminating $\mathbf{H}$. Introduction of the wave number $k = 2 \pi f \sqrt{\mu_0\epsilon_0}$ yields
\begin{equation}\label{eq:Maxwell-eig-cont}
\begin{aligned}
\nabla\times{\left(\nabla\times{\mathbf{E}}\right)} &= k^2 \mathbf{E} && \text{in }\Omega(\boldsymbol{Y})\\
\mathbf{E}\times\mathbf{n} &= 0 && \text{on }\partial\Omega(\boldsymbol{Y}),
\end{aligned}
\end{equation}
for $\mathbf{E} \neq 0$ and $\nabla\cdot{\mathbf{E}}=0$. One should be aware  that, although not explicitly specified, the field $\mathbf E$ and the eigenfrequencies $f$ depend on the shape parameters $\boldsymbol{Y}$. For each mode, i.e., a solution of \eqref{eq:Maxwell-eig-cont} in the passband, there is a phase shift between fields of neighboring cells  that can vary from 0 to $\pi$ radians. The $\pi$-mode, with frequency $f_\pi$, of the TM$_{010}$ passband is used in multi-cell cavities for acceleration. 

In order to maximize the voltage across the cavity, the length of the middle-cells $L$ is fixed as $L=\beta \lambda/2$, where $\lambda$ refers to the wavelength of the $\pi$-mode, see \cite{Padamsee_2015aa}. Thus as the particle traverses a cell, the direction of the field reverses such that the particle is exposed to an identically directed electric field along the whole multi-cell cavity. 

The modes in a passband have a small frequency difference. If the frequencies of the modes in the fundamental passband are very close to each other, there is a risk of exciting a mode close to $f_\pi$ by the RF generator. 

The spread of modes in the first passband is reflected in the cell-to-cell coupling coefficient, which is defined as~\cite{Belomestnykh2005}
\begin{equation}
k_{\text{cc}} = 2\frac{f_\pi - f_0}{f_\pi + f_0}, 
\label{eq:coupling}
\end{equation}
where $f_0$ refers to the lowest frequency in the passband. The cell-to-cell coupling coefficient is dimensionless and a sensitive quantity in the design phase. A large aperture radius ($R_\text{ir}$) typically gives rise to a stronger cell-to-cell coupling.

If the energy of the fundamental mode is evenly distributed in the cells, the accelerating voltage is maximized~\cite[p.129]{Padamsee1998}. Furthermore, a uniform field distribution allows for higher field magnitudes before reaching the surface electromagnetic (EM) field limit. The field flatness is a central figure of merit that indicates the uniformity of the field distribution of the fundamental mode between the cells. In this paper, the field flatness is defined as
\begin{equation}
\mathcal{F}=\frac{\min_{i=1,\ldots,N_\text{c}} |E_{\text{ax,max}}^{(i)}|}{\max_{i=1,\ldots,N_\text{c}} |E_{\text{ax,max}}^{(i)}|}  
\label{eq:fieldflatness}
\end{equation}
where $E_{\text{ax,max}}^{(i)}$ refers to the maximum axial electric field in cell $i$. 

The resonant frequencies strongly depend on the geometry parameter in each cell, i.e., the equatorial radii $R_{\text{eq}}^{(i)}$, with $i=1,\ldots,N_\text{c}$, and the iris radii $R_{\text{ir}}^{(j)}$, with $j=1,\ldots,N_\text{c}+1$. In Section \ref{sec:uncertainty} we consider perturbations of the form $R_{\text{eq}}^{(i)} + \Delta R_{\text{eq}}^{(i)}$ and $R_\text{ir}^{(j)} + \Delta R_\text{ir}^{(j)}$ for the equatorial radii and the iris radius, respectively. Then, 
the parameter vector is given as
\begin{equation}\boldsymbol{Y} =  [\Delta  R_\text{eq}^{(1)}, \ldots, \Delta  R_\text{eq}^{(9)},\Delta R^{(1)}_\text{ir},\ldots, \Delta R_\text{ir}^{(10)}].\label{eq:paravec}\end{equation} These perturbations change the resonant frequency of the respective cell(s) and consequently affect the frequency and the field distribution of the $\pi$-mode in the multi-cell cavity. It has been observed, that for the $\pi$-mode, the change in the field amplitude of each cell is proportional to the frequency change by a factor of $\propto 1/k_{\text{cc}}$~\cite{Sekutowicz1999,Nagle1967}. Thus, a small cell-to-cell coupling increases the sensitivity of the field profile with respect to geometrical perturbations.

\section{Cavity Manufacturing}
\label{sec:manufacturing}

The European X-ray Free Electron Laser \cite{XFEL_report} facility is constructed to produce X-ray pulses with the properties of laser light and at intensities much brighter than those produced by conventional synchrotron light sources. The superconducting linear accelerator of the EXFEL has a length of almost 2.1 km and brings electrons to an energy of up to \SI{17.5}{GeV}. This is achieved by using a total of $808$ superconducting cavities installed in the three main linac sections and the injector. The production of $N_\text{cav}>808$ cavities \cite{Singer2016}, the largest in the history of cavity production, was realized by the two companies Research Instruments GmbH (RI) and Ettore Zanon S.p.A. (EZ). EXFEL uses nine-cell TESLA cavities build from solid niobium with a nominal $f_\pi=1,300$ MHz. Each cavity (see FIG.~\ref{fig:xfel_cavity}) consists of 10 main sub-components, welded together at the equator area. The sub-components consist of $8$ dumb-bells (DB) and $2$ end-groups (EGL, EGS), referring to short and long end-groups, respectively. A different shape of the end-groups' half-cells provides the desired asymmetry of the Higher Order Mode (HOM) field distributions and increases the efficiency of their extraction.
\begin{figure*}[t!]
\centering
\includegraphics{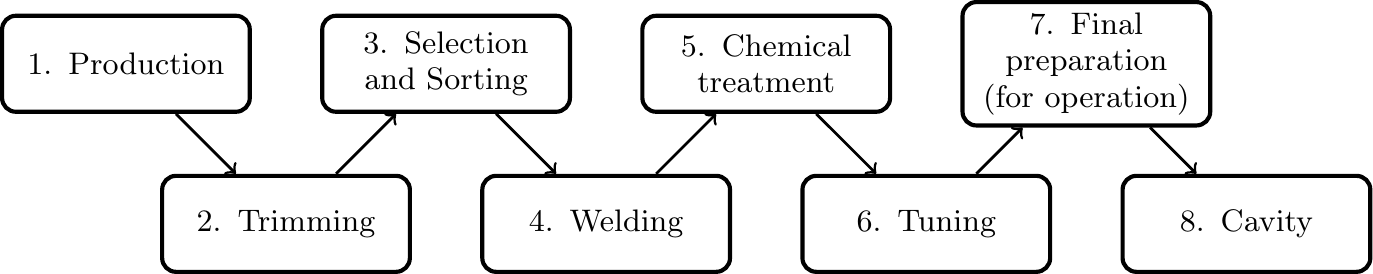}
	\caption{Different steps of the TESLA cavity production chain.}
	\label{fig:production_chain}
\end{figure*}	

Geometric deviations of the inner cavity shape, the cavity length, and the spectra of frequencies as well as deviations in HOM field distributions occur due to random inaccuracies during manufacturing. These uncertainties have a strong impact on the quantities of interest described in Section \ref{sec:quantities_of_interest}. 
Hence, dedicated measures are undertaken during production to 
ensure acceptable tolerances according to the EXFEL cavity specification, in particular to obtain $\mathcal F>90\%$ and to keep the deviation of $f_\pi$ below $\SI{100}{kHz}$. 
These measures, together with sources of uncertainty, are summarized in FIG.~\ref{fig:production_chain} and described in detail in the following.

\begin{enumerate}
		\item \textbf{Step:} Production\\
In this step $8 N_\text{cav}$ DBs, consisting of $2$ half cells each, $N_\text{cav}$  EGSs and $N_\text{cav}$ EGLs are produced. 
		\item \textbf{Step:} Trimming\\
The target of this step is to compensate for shape deviations by trimming the components. It is applied to all components (DB, EGS and EGL) and allows to obtain the necessary cavity length and frequencies with required accuracy. 
		\item \textbf{Step:} Selection and Sorting\\
		The manufacturer selects two end groups and $8$ dumb-bells. To minimize the influence of shape variations for the eight DBs on the asymmetry of the HOM field distribution, the DBs are sorted during the cavity assembly: a DB with average frequency is installed at the last position (position 8), the remaining DBs are installed in order of decreasing frequency (see FIG.~\ref{fig:sorted_freqs}).
\begin{figure}[t!]
\includegraphics{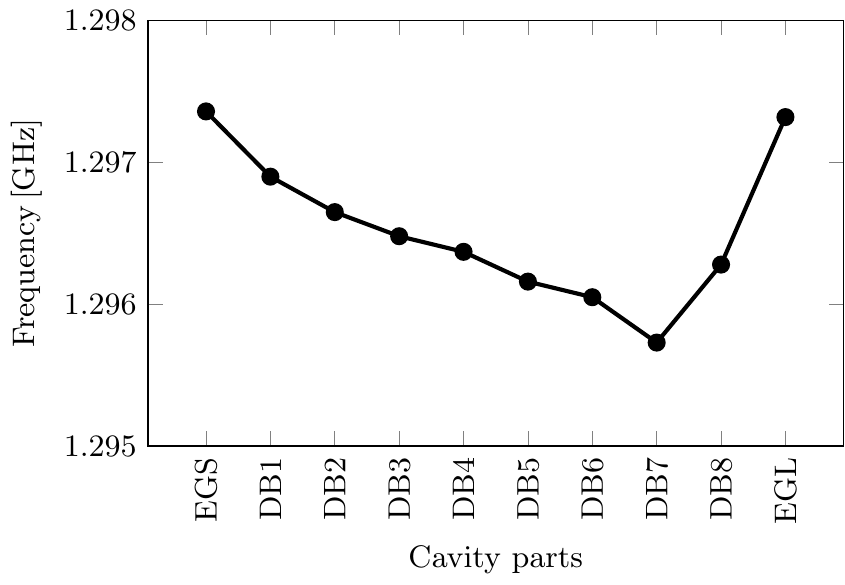}
\caption{Eigenfrequencies of dumb-bells and end-groups for an exemplary cavity from the DESY database \cite{desy_db}.}
\label{fig:sorted_freqs}
\end{figure}

		\item \textbf{Step:} Welding\\
All components (EGL, DBs, EGS) are welded to each other, which induces shape deformations. The materials from the different suppliers exhibit different shrinkages at the welding joint. The resulting equator diameters may be slightly different depending on the cavity position during welding, see \cite{Sulimov2015}.
		\item \textbf{Step:} Chemical treatment\\
The chemical treatment removes impurities and spikes, see \cite{Singer2016}, \cite{Sulimov2016}. The homogeneity of the removed material from the cavity surface depends on many parameters of the process and the facility. The electrochemical polishing treatment is usually less homogeneous and more unstable than the equator welding.
		\item \textbf{Step:} Tuning\\
The cavity is tuned, i.e., mechanically stretched or compressed, according to the procedure described by \cite{Kreps_1996aa} which adjusts $f_\pi$ with an accuracy of $\pm\SI{50}{kHz}$ and ensures $\mathcal{F}>98\%$ for the field flatness.
		\item \textbf{Step:} Final preparation (for operation)\\
The procedures applied in this step vary for different manufacturers and are shown in detail in FIG.~2 of \cite{Singer2016}. Those procedures include, e.g., final buffered chemical polishing etching, the integration of the cavities into the Helium tanks, a pressure test using water under the pressure of $6$ bar and the cool down to \SI{2}{K}.  
		\item \textbf{Step:} Cavity\\
All $N_\text{cav}$ cavities are operational and the statistics of the fundamental mode spectra are measured under \SI{2}{K}. A description of the measurement procedures is given below. 
	\end{enumerate}

Quality assurance by mechanical measurements of the inner surface dimensions becomes impossible after cavity welding and polishing. Measurement data can only be obtained by ultrasonic or RF measurements. These methods are used for control of equator welding stabilities \cite{Sulimov2015} or homogeneity of the polishing process \cite{Sulimov2016}. In the following the fundamental mode spectra are discussed for both manufacturers (RI and EZ) based on RF measurements. 

\begin{figure}[t!]
\includegraphics{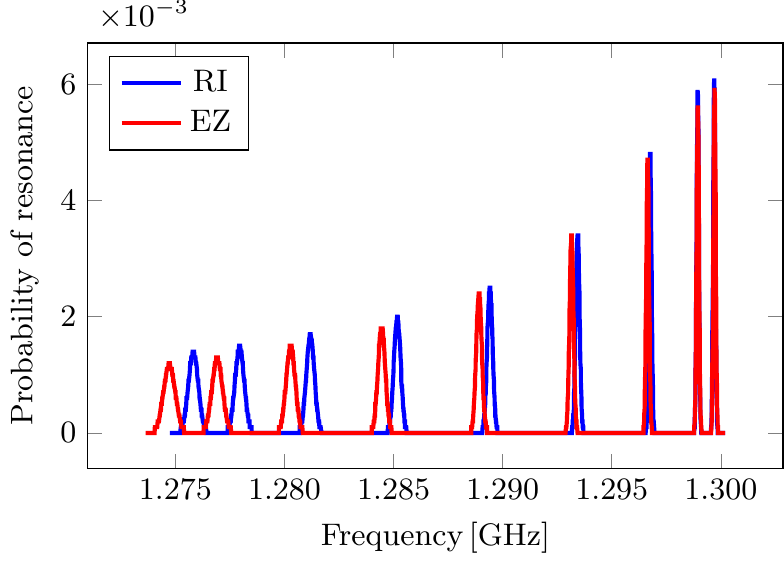}
\caption{Distribution of TM$_{010}$ spectra for different manufacturers of the European XFEL cavities, data from \cite{desy_db}.}
\label{fig:spectra_dist}
\end{figure}

\begin{figure}[t!]
\includegraphics{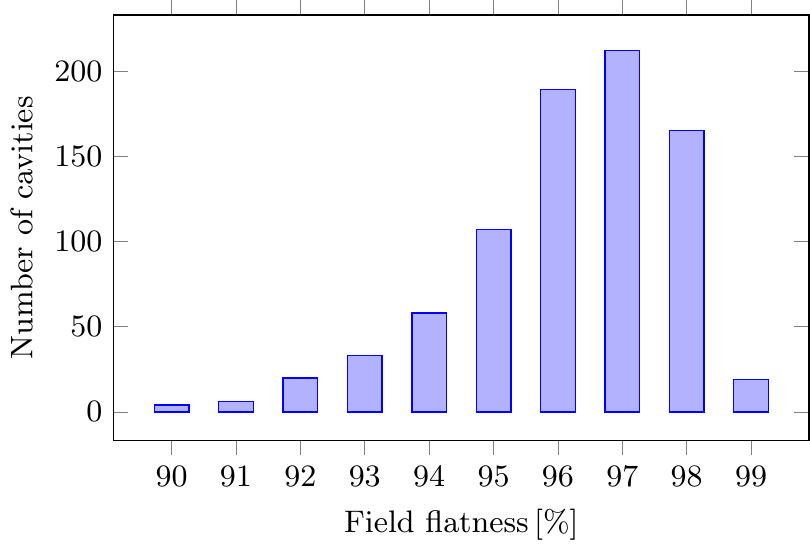}
\caption{Cavities with different field flatness, data taken from \cite{desy_db}.}
\label{fig:FF_stats}
\end{figure}

The measurements were carried out at operation temperature, i.e., cryogenic tests at 2\,\si{K} were used (FIG.~\ref{fig:spectra_dist}). The measurement results are presented in Table~\ref{tab:spectra_stats}. Note, that the $\pi-$mode finally operates at 1.3\,\si{GHz}, which is ensured by an additional tuning step (not described here). The results show that the standard deviation of the $\pi-$mode (mode 8) in the relaxed condition after cool down, is about 50\,\si{kHz}. This value increases with decreasing mode number. A similar behavior can be found for the differences 
$$\Delta f_m=\bigl|\mathbb{E}\bigl[f_{\text{RI},m}\bigr]-\mathbb{E}\bigl[f_{\text{EZ},m}\bigr]\bigr|$$
of the averaged frequencies for each mode $m$ between the two manufacturers $\textrm{RI}$ and $\textrm{EZ}$, which also increases with decreasing mode number, see Tab.~\ref{tab:spectra_stats}. In particular, the difference of average frequencies is about 14\,\si{kHz} for mode 8 and 1098\,\si{kHz} for mode 0, respectively. These results indicate a high deviation of the cell-to-cell coupling coefficient between the manufacturers with 
\begin{align*}
k_{\text{cc}} &= (1.854 \pm 0.016)\,\% \text{ (RI) } \text{and} \\
k_{\text{cc}} &= (1.941 \pm 0.021)\,\% \text{ (EZ)},
\end{align*}
which are given in the form of average value plus/minus one standard deviation as obtained from database, see also Tab.~\ref{tab:irisDeformations}. Preliminary studies \cite{Sulimov2013} show that only variations of the iris radius can explain such high deviation in the cell-to-cell coupling coefficient. This will be analysed in more detail numerically in Section \ref{sec:uncertainty}. Finally, the field flatness distribution for the EXFEL cavities is presented in FIG.~\ref{fig:FF_stats}. It can be observed that at least $70\%$ of the cavities possess a field flatness of more than $95\%$.

\begin{table}[t!]
\renewcommand{\arraystretch}{1.2}
\caption{TM$_{010}$ spectra for the European XFEL cavities, data taken from \cite{desy_db}.}
\label{tab:spectra_stats}
\begin{tabular}{|c|c|c|c|c|}
    \hline
    \multirow{3}{*}{Mode} & \multicolumn{4}{c|}{Frequency, MHz}\\\cline{2-5}
    &\multicolumn{2}{c|}{Average}&\multicolumn{2}{c|}{St Dev}\\\cline{2-5}
    & RI & EZ & RI & EZ\\\hline
    0 & 1275.832 & 1274.734 & 0.226 & 0.268\\\hline
    1 & 1277.951 & 1276.926 & 0.212 & 0.245\\\hline
	2 & 1281.194 & 1280.294 & 0.186 & 0.208\\\hline
	3 & 1285.178 & 1284.460 & 0.157 & 0.168\\\hline
    4 & 1289.423 & 1288.921 & 0.124 & 0.129\\\hline
    5 & 1293.442 & 1293.161 & 0.089 & 0.090\\\hline
    6 & 1296.762 & 1296.649 & 0.064 & 0.066\\\hline
	7 & 1298.944 & 1298.946 & 0.052 & 0.054\\\hline
	8 & 1299.702 & 1299.716 & 0.050 & 0.052\\\hline
\end{tabular}
\renewcommand{\arraystretch}{1.0}
\end{table}

\section{Uncertainty quantification}
\label{sec:uncertainty}

A stochastic setting is adopted here to model manufacturing and measurement uncertainties and assess their influence on the cavity design, i.e. each component of $\boldsymbol{Y}$ becomes a random variable
\begin{equation}
\begin{split}\boldsymbol{Y}_\text{prod}(\theta) =  [\Delta  R_\text{eq}^1(\theta), \ldots, \Delta  R_\text{eq}^9(\theta),\\ \Delta R^{(1)}_\text{ir}(\theta),\ldots,R_\text{ir}^{(10)}(\theta)]\label{eq:paravec_random}\end{split}\end{equation}
where $\theta$ denotes a random outcome. 

Uncertainty quantification encompasses various methods for uncertainty propagation, Bayesian inverse problems, optimal experimental design and robust optimization, among others. The reader is referred to \cite{ghanem1991,xiu2010,lemaitre2010} for a detailed background. In this work, uncertainty propagation is of great interest, in particular, propagating distributions of cell deformation parameters to distributions of quantities of interest, such as the cell-to-cell coupling coefficient. This can be achieved by sampling according to the underlying distribution and repetitively solving the cavity eigenvalue problem. Thereby, surrogate modeling is a key step to keep the computational workload manageable. The probability distributions of the cavity geometry parameters are modeled based on descriptions of the manufacturing process, as described in Section \ref{sec:manufacturing}. A more general approach would consist in inferring these input distributions from measured RF data, which would require the solution of an inverse cavity eigenvalue problem. Such a study is out of the scope of the present work and can only be briefly addressed here, applying significant simplifications. In fact, the present study, should be considered as a step towards a more complete treatment of uncertainties in the cavity eigenvalue problem.

To solve Maxwell's equations we use the variational formulation of~\eqref{eq:Maxwell-eig-cont}. It reads: 
find $k$ and $\ensuremath{\mathbf{E}}$ such that
\begin{equation}\label{eq:Maxwell-eig-var}
 \left(\ensuremath{\nabla \times \ensuremath{\mathbf{E}}},\ensuremath{\nabla \times \ensuremath{\mathbf{v}}}\right) 
  = k^2 \left(\ensuremath{\mathbf{E}},\ensuremath{\mathbf{v}}\right) \quad \forall \ensuremath{\mathbf{v}}, 
\end{equation}
where we assume that $\mathbf{E}$ and $\mathbf{v}$ are from the functions space of square-integrable vector fields with square-integrable \textbf{curl}. For further information on function spaces in the
context of Maxwell's equations, the reader is referred to \cite{Monk_2003aa}.
In order to numerically solve~\eqref{eq:Maxwell-eig-var} we 
use a finite-dimensional spaces such that
\begin{equation}\label{eq:Maxwell-eig-var-discrete}
\left(\ensuremath{\nabla \times \ensuremath{\mathbf{E}}_h},\ensuremath{\nabla \times \ensuremath{\mathbf{v}_j}}\right) 
  = \ensuremath{k_h}^2 \left(\ensuremath{\mathbf{E}}_h,\ensuremath{\mathbf{v}_j}\right) \quad \forall \ensuremath{\mathbf{v}_j}.
\end{equation}
with 
\begin{equation}
\ensuremath{\mathbf{E}}_h = \sum_{j=1}^{\ensuremath{N_{\text{dof}}}} e_j \ensuremath{\mathbf{v}_{j}}.
\end{equation}
where $N_{\text{dof}}$ denotes the number of degrees of freedom.

In this paper, calculations are carried out with the help of \textsc{Superlans} code~\cite{SLANS}. \textsc{Superlans} is a 2D-axisymmetric finite-element-based code used for the calculation of the monopole modes of azimuthally symmetric geometries. For each simulation, the data describing the contour of the cavity is created by \textsc{Matlab}~\cite{Matlab} and saved in a format readable by \textsc{Superlans}. \textsc{Superlans} is then called by \textsc{Matlab} to triangulate the geometry, solve the resulting eigenvalue problem and calculate the relevant secondary parameters. The results are finally read by \textsc{Matlab} for the post-processing. The detailed description of the uncertainty modeling steps, which are the numerical counterpart of the production chain in Section \ref{sec:manufacturing}, are as following:
 
\begin{enumerate}
\item \textbf{Step:} Production\\
We generate $\tilde N_\text{cav} =10^6$  random ("virtual") cavities by creating, in turn, seven independent random mid cells and two end cells per cavity. Since  In  particular, those virtual cavities are obtained by drawing random numbers for the vector $\boldsymbol{Y}_\text{prod}$ defined in \eqref{eq:paravec_random} where we assume independence of the random variables. Note that this assumption can often be justified for mass-production processes, in particular, as the individual dumb-bells are produced independently of each other. Although, the normal distribution seems to be the right choice, we opt for beta distribution in the  range \SIrange{-0.3}{0.3}{mm}. Beta distributions can be used to approximate normal distributions but have bounded support \cite[Appendix B]{xiu2010}, which is very important in numerical studies to avoid non-physical parameter configurations. FIG.~\ref{fig:beta_pdf} represents such an approximation with probability density function (PDF)
\begin{align*}
\varrho(y) = \frac{140}{(u-l)^7}\begin{cases} (y-l)^3(u-y)^3, &l<y<u,\\0,  &\text{else},
\end{cases} \label{eq:beta44_pdf}
\end{align*}
where $l=\SI{-0.3}{mm}$ and $u=\SI{0.3}{mm}$ denote the lower and upper bound, respectively. In this particular case the shape parameters of the beta distribution were chosen such that a normal distribution with a $2\sigma$ interval of \SIrange{-0.2}{0.2}{mm} is approximated, as illustrated in FIG.~\ref{fig:beta_pdf}.
\begin{figure}
	\centering
	\includegraphics{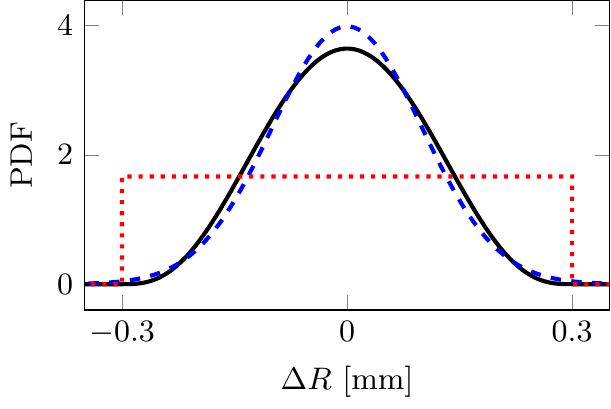}
		\caption{Black: PDF of beta distributed radius variation with support in $\left[\SI{-0.3}{mm}, \SI{0.3}{mm}\right]$. Blue, dashed: PDF of normal distribution with $\mu = \SI{0}{mm}$ and $\sigma = \frac{0.2}{2}\,\mathrm{mm}$. Red, dotted: PDF of uniform distribution with support in $\left[\SI{-0.3}{mm}, \SI{0.3}{mm}\right]$.}
	\label{fig:beta_pdf}
\end{figure}
We note that the large number of virtual cavities $\tilde N_\text{cav}$ does not lead to prohibitive computational cost, as surrogate modeling is employed which will be explained later.

\item \textbf{Step:} Trimming\\
In contrast to real manufacturing, in the simulation approach, we build cavities out  of elementary cells instead of dumb-bells. In this setting, it is not clear how to explicitly model the trimming, however it is implicitly taken into account by the cavity length limitation \eqref{eq:length_constraint} which is considered in the next step. 
\item \textbf{Step:} Selection and Sorting\\
The virtual 9-cell cavity is obtained from the random building blocks of step 1 as follows: We apply a sorting procedure which computes for all middle cells the fundamental resonance frequency (by solving a one-cell eigenvalue problem), places the cell closest to the average frequency at the 8$^\textrm{th}$ position and orders the remaining cells according to decreasing frequency between positions 2 and 7. 
Additionally, a total length constraint 
\begin{equation}\sum_{i=1}^{N_\text{c}} \Delta L^{(i)}<\SI{3}{mm}\label{eq:length_constraint} \end{equation}is enforced. If the constraint is violated after tuning (step 6), the corresponding virtual cavity is disregarded. In real manufacturing, the constraint is already incorporated by trimming and compensation effects.

\begin{figure*}
	\includegraphics{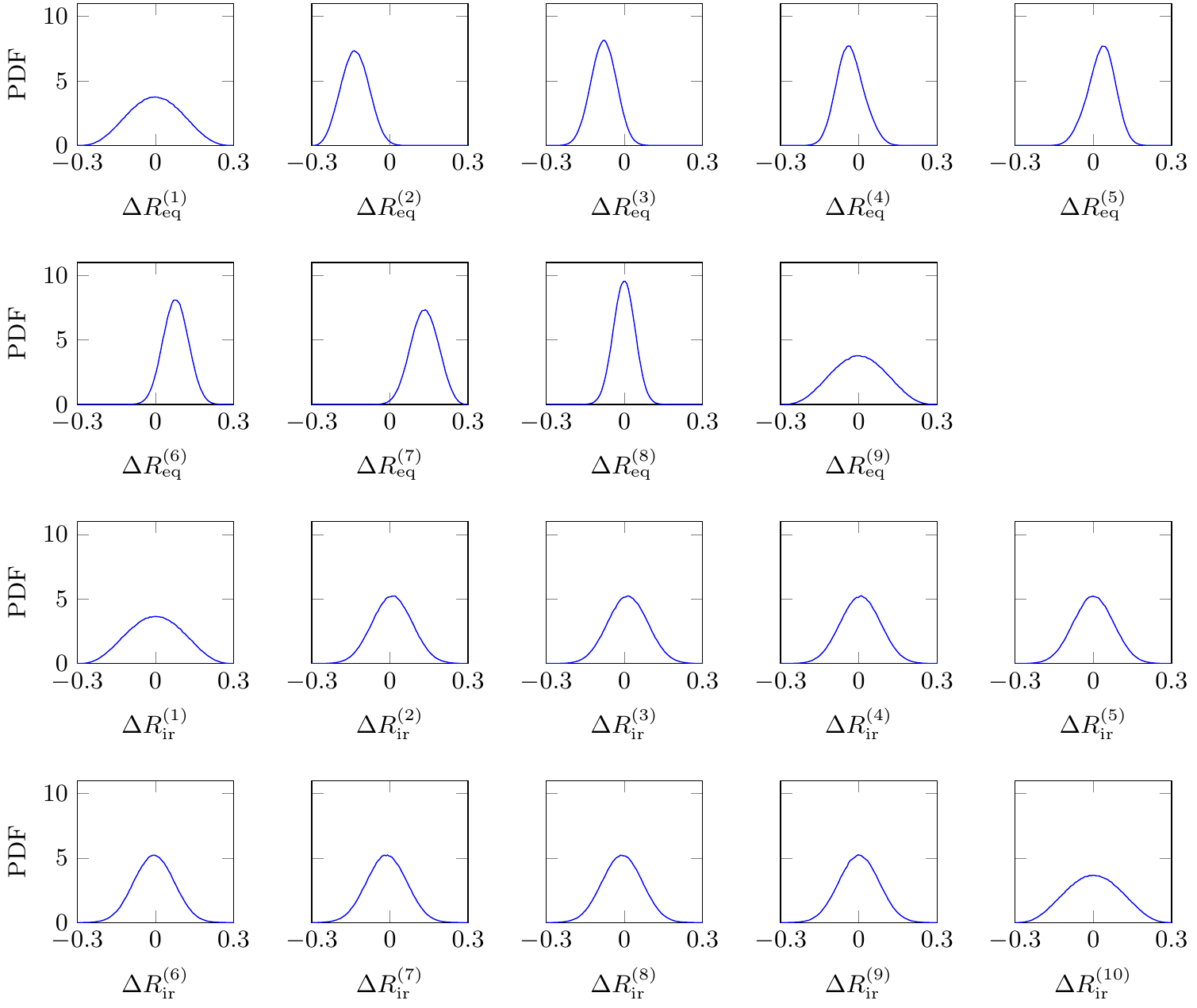}
		\caption{Kernel density estimates of the 19 correlated input random variables.}
	\label{fig:1D_KDEs}
\end{figure*}

\item \textbf{Step:} Welding\\
Since we consider cells instead of dumb-bells for the simulation, the welding is modeled by averaging the iris radii of adjacent cells. Note that the impact of eccentric deformations by (possible) miss-alignment of cells  has been investigated in \cite{Georg_2019aa}.  It was found that this type of deformation has negligible impact on the fundamental mode spectrum and is hence omitted here.

The steps of selection, sorting and welding influence the probability distribution. Hence, we introduce a new random vector  $\boldsymbol{Y}_\text{sort}(\theta)$ where the density is estimated with kernel density techniques. Kernel density estimation is a non-parametric technique to infer a continuous probability density function from a sample. Here, we employ an Epanechnikov kernel \cite{Epanechnikov} on the selected (and sorted) sample $\{\boldsymbol{Y}^{(m)}_\text{sort}\}_{m=1}^{\tilde N_\text{sel}}$ of size $\tilde N_\text{sel}=809641$, which complies with the length constraint. The estimated densities are presented in Fig.~\ref{fig:1D_KDEs}. 
		\item \textbf{Step:} Chemical treatment\\
An appropriate modeling of chemical treatment would require very fine resolutions or even multi-scale analyses which cannot be carried out in the present setting, see also \cite{Sulimov2016} for further information. 
\item \textbf{Step:} Tuning\\
The virtual tuning procedure we apply, considers each cell individually and is therefore unable to incorporate field flatness constraints directly. However, we  observe that  an acceptable field flatness of at least $96\%$ was obtained for all eigenvalue problems that are eventually solved. 
In particular, each cell is tuned to \SI{1.3}{GHz} by changing its length $L^{(i)}$. Computationally, this requires the solution of a non-linear root finding problem for the objective function
\begin{equation*}
f_{\text{obj}}^{(i)}\bigl(L^{(i)}\bigr) = f_0^{(i)}\bigl(L^{(i)}\bigr) - \SI{1.3}{\giga\hertz},
\end{equation*}
where $f_0^{(i)}$ denotes the fundamental eigenfrequency of cell $i$ (one-cell eigenvalue problem). The root finding problem is solved using \texttt{fzero} in \textsc{Matlab}. However, we note that different root finding schemes, e.g., Newton's method or bisection, could be employed as well.

In FIG.~\ref{fig:test_tuning_smoothness} the values of the 9-cell cavity accelerating frequency are depicted for different choices of the length and equatorial radius of the first cell.  Considering the \SI{1.3}{\giga\hertz} contour (in black in the figure), we observe that the tuning process correctly identifies a length such that a value of $\SI{1.3}{GHz}$ is obtained (the magenta dots in the figure lie on the contour line as expected).  
\begin{figure}[h]
\includegraphics{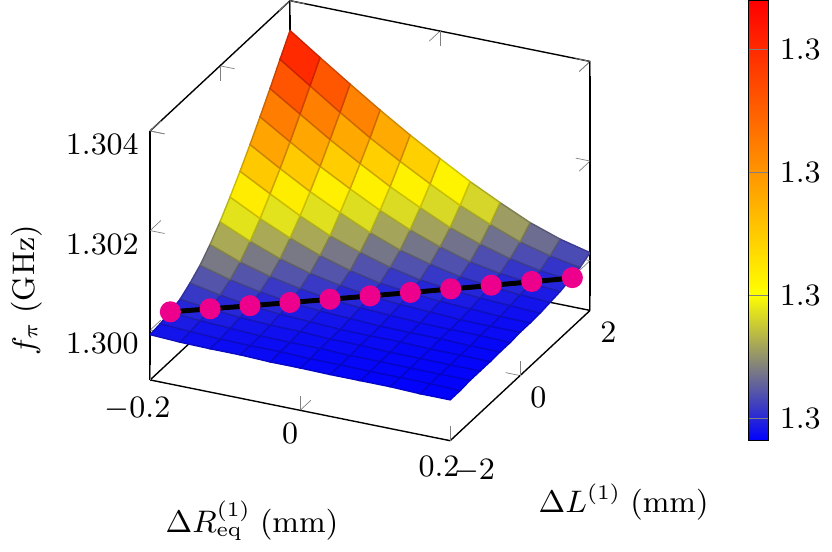}
\caption{The surface represents the accelerating frequency of the 9-cell cavity for different values of the changes in the equatorial radius  $ \Delta R_\text{eq}^{(1)} $ and in the length $\Delta L^{(1)} $ of cell 1. The black line is the \SI{1.3}{\giga\hertz} contour line. The magenta points are the tuning values for $ \Delta {L^{(1)}} $ obtained for a given value of $ \Delta R_\text{eq}^{(1)} $. }\label{fig:test_tuning_smoothness}
\end{figure}
		\item \textbf{Step:} Final preparation (for operation)\\
Due to insufficient data and involved numerical modeling this step cannot be carried out in the setting employed in this work.

\item \textbf{Step:} Cavity\\
The Maxwell eigenvalue problem is solved to obtain the first nine modes for each virtual cavity and statistics of the spectra are computed. It shall be noted that, in general, eigenvalue tracking \cite{Georg_2019aa} should be employed to ensure a consistent matching of the eigenfrequencies. However, it has been observed that, in this case, the fundamental eigenfrequencies do not cross with respect to parameter changes for the considered variations. Hence, mode tracking is not applied here. The flow diagram of simulation steps is given in Fig.~\ref{fig:SimulationLayout}. 
	\end{enumerate}
 
  \begin{figure*}[!ht]
	\includegraphics[width=2.0\columnwidth]{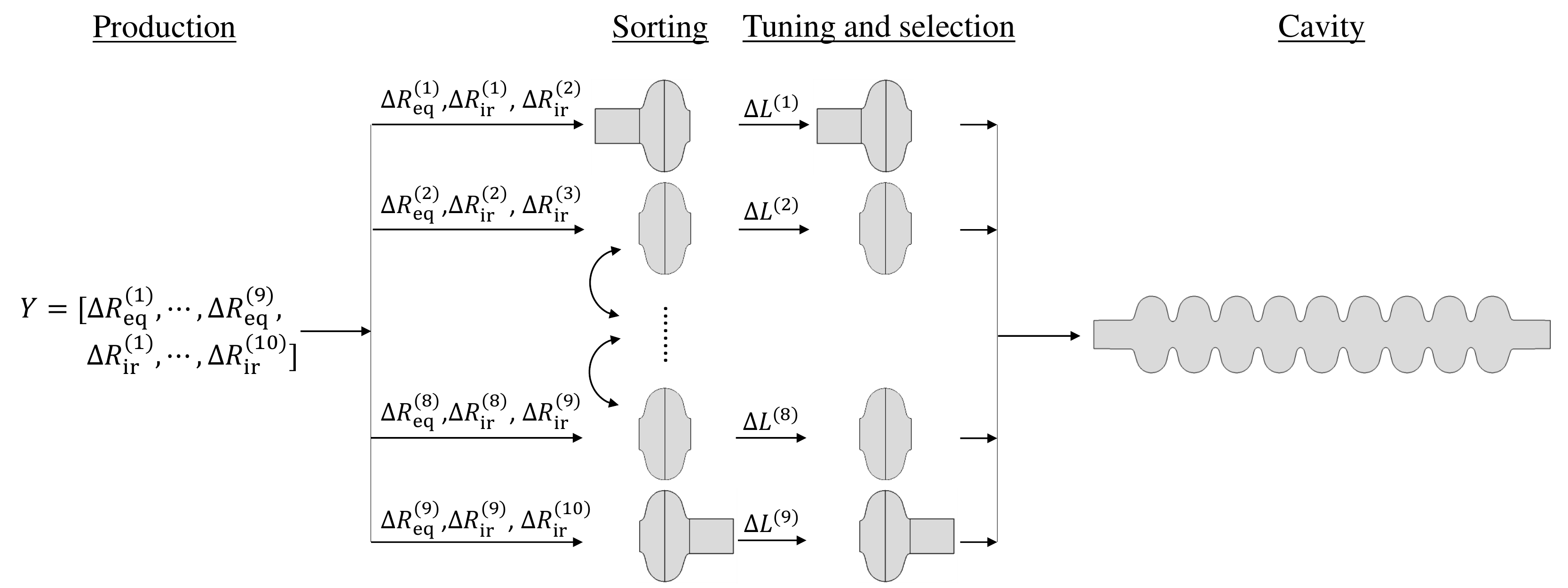}
	\caption{The flow diagram of simulations steps.}
	\label{fig:SimulationLayout}
\end{figure*}

In order to avoid the tremendous computational cost of repeatedly solving a large number of eigenvalue problems in step 3 and step 8, surrogate modeling is employed. In this work a dimension-adaptive interpolation  scheme \cite{narayan2014}  is used to  construct   accurate  polynomial  surrogate  models. In particular, the  algorithm  described  in  \cite[Chapter  3.2]{georg2018}  is adapted  to address  the case  of  multiple  quantities  of  interest. This algorithm  constructs polynomial models adaptively which yields high computational efficiency. Moreover, we control the associated approximation errors by cross-validation to be sufficiently small. For convenience of the reader, we recall the main ideas of the employed algorithm in appendix~\ref{sec:surrogate} and refer for details to \cite{georg2018}.

The 4-variate polynomial surrogate models employed in step 3 are computed by solving 50 one-cell eigenvalue problems while 500 evaluations of the 9-cell eigenvalue problem are used to construct the 19-variate polynomial surrogate model for step 8. For all surrogate models, cross-validation  indicates an error below $\SI{10}{kHz}$ for all fundamental resonance frequencies which is smaller than the standard expected deviations.

\begin{figure}
	\centering
	\includegraphics{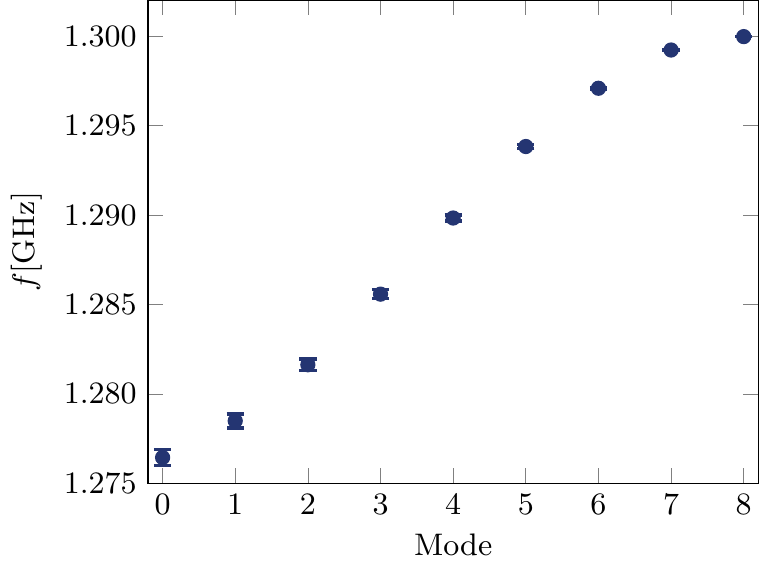}
		\caption{Mean values of the frequencies of the first passband with bars indicating the $ 3\sigma $ deviation intervals. Results are computed with the adaptive Leja sparse grid algorithm and 500 function evaluations.}
	\label{fig:uq_freqs}
\end{figure}
\begin{table}
\renewcommand{\arraystretch}{1.2}
\caption{Mean values and standard deviations of the first passband in the tuned configuration.}
\label{tab:uq_freqs}
\pgfplotstabletypeset[
	col sep=comma,
	columns/mode/.style={column name=~Mode~, column type={|c}, precision=0},
    columns/mean/.style={column name=~Mean\,[MHz]~~, column type={|c}, precision=2, fixed zerofill},
    columns/ThreeSigma/.style={column name=~Std. dev.\,[MHz]~~, column type={|c|}, preproc/expr={##1/3}, fixed, precision = 2, fixed zerofill},
    every head row/.style={before row=\hline,after row=\hline},
    every last row/.style={after row=\hline},
    ]{Statistics_mode_spectrum809641Samples_19_Params.csv}
\renewcommand{\arraystretch}{1.0}
\end{table}

Now, all statistical quantities of interest can be obtained from the sample computed in step 8. In particular, we employ {nbiased statistical estimates for expectation and standard deviation, which are depicted in FIG.~\ref{fig:uq_freqs} and Table~\ref{tab:uq_freqs}. In the same way, we estimate expectation and standard deviation of the cell-to-cell coupling coefficient $k_\text{cc}$ 
\begin{align}
\mathbb E[k_\text{cc}] \approx 1.82802, \quad 
\operatorname{Std}[k_\text{cc}] \approx 0.01897.
\end{align}
In the following, we employ Sobol indices as sensitivity measure, despite the fact that they are defined for independent parameters.
We use the \textsc{OpenTurns} implementation of Saltelli's algorithm \cite{saltelli2002} to estimate Sobol indices of the cell-to-cell coupling coefficient for the $10$ input parameters $\boldsymbol{Y}$. Readers interested in details on this algorithm are referred to appendix~\ref{app:Sobol}. The results presented in FIG.~\ref{fig:sobol_Kcc_19params} (top) are obtained by evaluating $4\cdot 10^7$ times the surrogate model. It can be observed that deviations in the middle cells, in particular of the respective iris radii, have significantly larger contributions to the cell-to-cell coupling coefficient. Additionally, we estimate Borgonovo indices, which represent another global sensitivity index family taking correlations in the input data into account. To this end, we employ an approach based on kernel density estimation, see for instance \cite{da2015global} and references therein, on the sample $\{\boldsymbol{Y}_\text{sort}^{(m)}\}_{i=1}^{\tilde N_\text{sel}}$ and the corresponding surrogate model evaluations of $k_\text{cc}$. Contrary to Sobol indices which decompose the output variance with regard to different model input contributions, in the Borgonovo approach, the dependency between input and output parameters is analysed, see Appendix~\ref{app:Sobol} for more details. Hence, different sensitivity magnitudes can be expected in each case. However, also in the case of Borgonovo sensitivity analysis, similar results are obtained. In particular, the variations in the middle cells can be identified to have a larger impact than the corresponding end-cell variations.
The surprisingly large contribution $\Delta R_\text{eq}^{(8)}$, in comparison to the variance-based analysis, can most probably be explained by the special treatment of cell $8$ in the sorting procedure. Since we place the cell closest to the average frequency at position $8$, a large variation of $\Delta R_\text{eq}^8$ might implicitly also imply larger variations of the other middle cells. To confirm this interpretation, we additionally repeated this study with a different sorting procedure which omits the special treatment of cell $8$. In this case, the sensitivity of $\Delta R_\text{eq}^{(8)}$ was reduced drastically, as expected.

As in \cite{Sulimov2013}, we proceed by making the simplifying assumption,  that, for the rest of the paper, all iris radii deviations $\Delta R_\text{ir}^{(i)},\,i=1,\ldots,N_\text{c}+1$ of a cavity are the same and accordingly drop the index $i$. This allows us to carry out a preliminary inverse analysis, i.e., to compute geometric variations from the cell-to-cell coupling coefficient. To this end, we first repeat the computation of sensitivity indices, by rerunning the full simulation workflow, where we now model the overall iris radii deviation $\Delta R_\text{ir}$ as one beta distributed random variable in the range of $\pm 0.3\,\text{nm}$. The corresponding Sobol indices, computed using $2.2\cdot 10^6$ surrogate model evaluations in Saltelli's algorithm, as well as the Borgonovo indices, are shown in Fig.~\ref{fig:sobol_Kcc_10params}.

As expected, the cell-to-cell coupling coefficient is heavily influenced by the iris radius while the equatorial radii have significantly less impact. In particular, the Sobol coefficients indicate that more than $95\%$ of the variance can be attributed to changes in the iris radius $\Delta R_\text{ir}$.

\begin{figure*}
	\begin{subfigure}{\textwidth}
\includegraphics{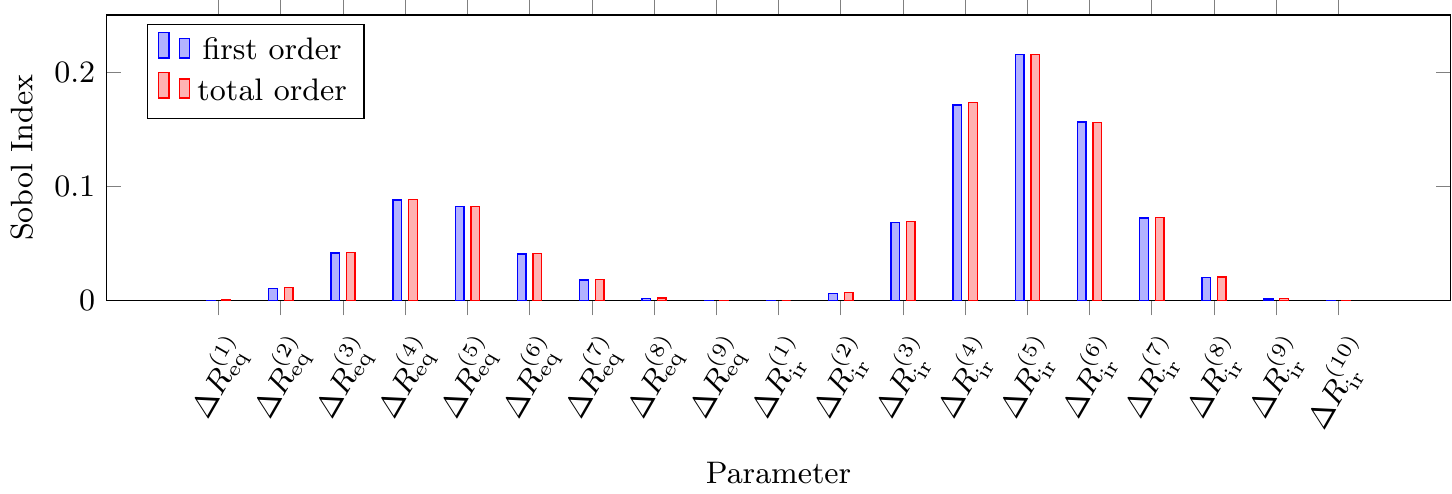}
\end{subfigure}
\begin{subfigure}{\textwidth}
\includegraphics{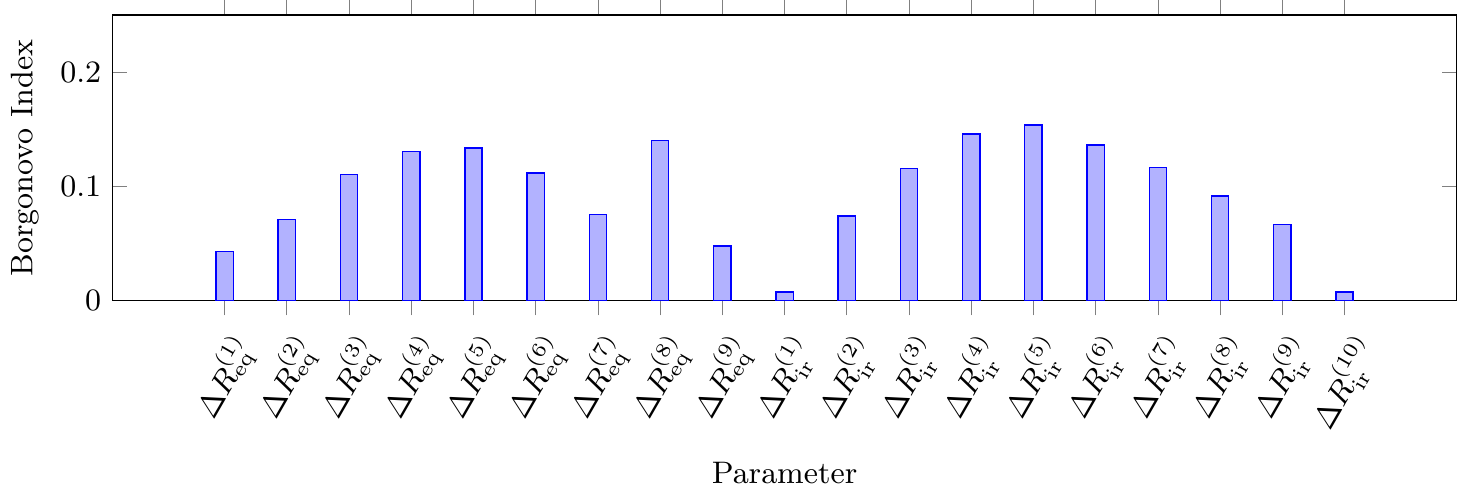}
\end{subfigure}
\caption{Sobol indices (top) and Borgonovo indices (bottom) for the cell-to-cell coupling $k_{\text{cc}}$.}\label{fig:sobol_Kcc_19params}
\end{figure*}

\begin{figure}
	\begin{subfigure}{\linewidth}
\includegraphics{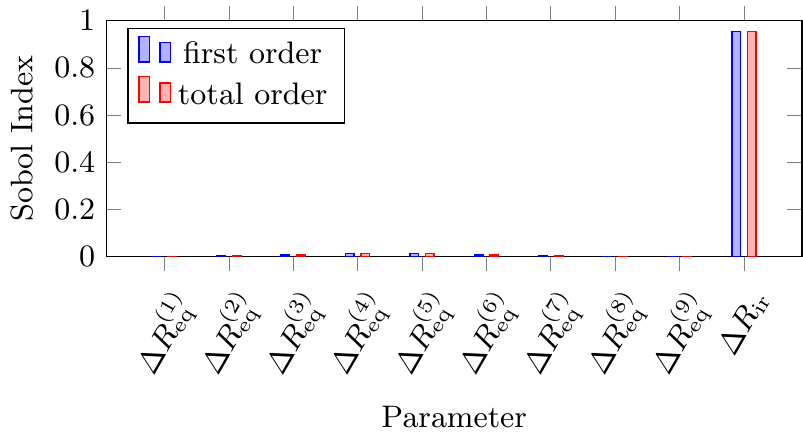}
\end{subfigure}
\begin{subfigure}{\linewidth}
\includegraphics{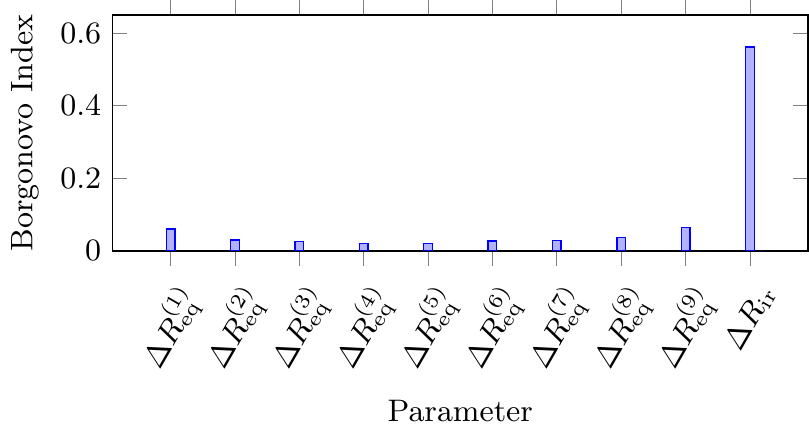}
\end{subfigure}

\caption{Sobol indices (top) and Borgonovo indices (bottom) for the cell-to-cell coupling $k_{\text{cc}}$, when all iris radii are deformed by the same value of $\Delta R_\text{ir}$.}\label{fig:sobol_Kcc_10params}
\end{figure}

\begin{figure}
	\centering
\includegraphics{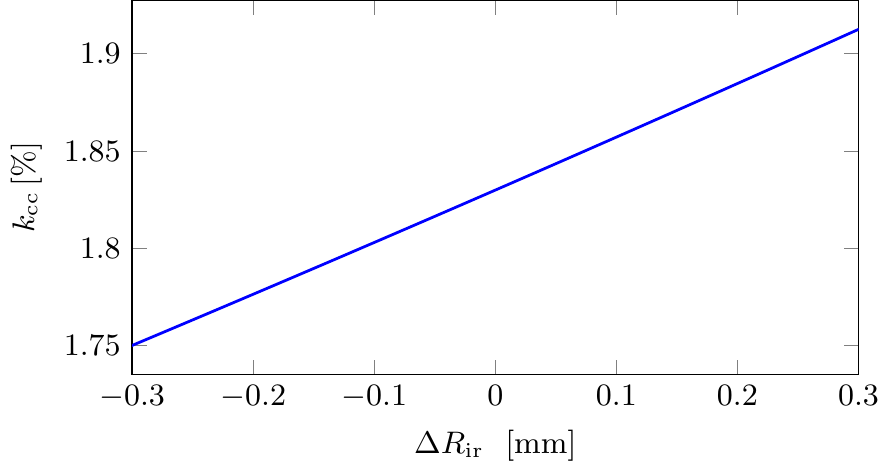}
\caption{Coupling coefficient $k_\text{cc}$ with respect to change in iris radius $R_\text{ir}$.}
\label{fig:Kcc_fun}
\end{figure}
To conduct a preliminary inverse analysis, based on the previous findings, we neglect the deformations in the equatorial radii in the following. We note that considering all parameters individually in an inverse study would be numerically much more challenging, as it would require approaches which assure that the problem remains well-posed \cite{stuart_2010}. Such a study is considered as out of the scope of the present work. In this case, the task consists in estimating the parameter $\Delta R_\text{ir}$ from measurements of the cell-to-cell coupling coefficient $k_\text{cc}$. FIG.~\ref{fig:Kcc_fun} depicts the associated relation which appears to be  monotonic and almost linear in the considered range. We collected measurements of the fundamental mode spectra for $N_\text{cav}=826$ cavities from the DESY Database \cite{desy_db}. Parameter estimation is then carried out by numerically inverting \begin{equation}\Delta R_{\text{ir},i}\mapsto k_{\text{cc},i},\,i=1,\ldots,N_\text{cav},\end{equation} where it should be noted that the  \textit{subscript} $i$ refers to the cavity number, while in previous parts of the paper the \textit{superscript} $i$ referred to the position of a \textit{local} iris radius variations. This is implemented by reformulating the root-finding problem as an optimization problem and applying the \textsc{scipy} implementation of the \textsc{L-BFGS-B} algorithm \cite{byrd1995, zhu1997}, i.e., a quasi-Newton method. However, we note that this is not a sensitive choice and other root finding schemes, e.g., bisection or Newton's method could also be employed. The sample statistics are presented in Table~\ref{tab:irisDeformations}. 

The XFEL specification requires deviations in the iris radii before welding to be below \SI{0.2}{mm} with respect to the nominal value. It is expected that the mean values and deviations of the iris radii change during the production chain, for example the chemical treatment is considered to have a significant influence. This offset is estimated by our numerical model to be \mbox{$\mathbb E[\Delta R_{\text{ir},i}]=\SI{0.243}{mm}$}, cf. last row of Table~\ref{tab:irisDeformations}. Despite this offset, the standard deviation $\operatorname{Std}[R_{\text{ir},i}]$ from both vendors are still within the limits of the specification ($\SI{0.17}{mm}<\SI{0.2}{mm}$) after the full production chain. When considering the vendors separately, then both production processes, standard deviations \SI{0.057}{mm} and \SI{0.073}{mm}, operate approximately at a three-sigma level.

\begin{table}
\caption{Sample mean and sample standard deviation of cell-to-cell coupling coefficient for $N_\text{cav} =826$ cavities as well as statistics about corresponding estimated iris deformations.}
\label{tab:irisDeformations}
\begin{tabular}{|c||c|c||c|c|}
	\hline
		Manufacturer&$\mathbb{E}\left[k_{\text{cc},i}\right]$& $\operatorname{Std}\left[k_{\text{cc},i}\right]$& $\mathbb{E}\left[\Delta R_{\text{ir},i}\right]$ & $\operatorname{Std}\left[\Delta R_{\text{ir},i}\right]$
		\\\hline
		RI & 1.854 & 0.016 & \SI{0.087}{mm} & \SI{0.057}{mm}\\\hline
		EZ & 1.941 & 0.021 & \SI{0.400}{mm} & \SI{0.073}{mm}\\\hline
		RI+EZ & 1.897 & 0.047 & \SI{0.243}{mm} & \SI{0.170}{mm}\\\hline
\end{tabular}
\end{table}

\section{Conclusion}
\label{sec:conclusion}
In this contribution the manufacturing chain of the EXFEL cavity was summarized and translated into a simulation workflow considering uncertainties. To analyze the sensitivities of the uncertain parameters, we propose an efficient adaptive surrogate modeling technique. The numerical study confirms the expert knowledge that the iris radius is the most critical parameter for the cell-to-cell coupling coefficient. Finally, the surrogate model is used to infer the sample mean and sample standard deviation of the iris radius variations from frequency measurements. For both manufacturers the obtained standard deviations are within the specification. 

The practically very relevant determination of statistics for all design parameters from given measurements is still ongoing research, as well as the investigation of global sensitivities for HOM. Incorporating the available measurement data for HOMs 

\begin{acknowledgments}
The authors would like to acknowledge the support by the DFG (German Research Foundation) in the framework of the Scientific Network SCHM 3127/1,2 "Uncertainty quantification techniques and stochastic models for superconducting radio frequency cavities" that provided the basis for this collaborative work. The work of J. Corno, N. Georg and S. Sch\"ops is supported by the Excellence Initiative of the German Federal and State Governments and the Graduate School of Computational Engineering at Technische Universitat Darmstadt. N. Georg's work is also funded by the DFG grant RO4937/1-1. The work of S. Gorgi Zadeh was/is supported by the German Federal Ministry for Research and Education BMBF under contracts 05H15HRRBA and 05H18HRRB1, respectively. J. Heller's work was supported by the BMBF under the contracts 05K13HR1 and 05K16HRA. J. Schultz and U. R\"omer acknowledge the funding by the Deutsche Forschungsgemeinschaft (DFG, German Research Foundation) under Germany´s Excellence Strategy -- EXC 2163/1- Sustainable and Energy Efficient Aviation -- Project-ID 390881007.
\end{acknowledgments}

\appendix{

\section{Global Sensitivity Analysis}
\label{app:Sobol}
Sensitivity analysis is a powerful tool in order to investigate the model behaviour and analyse the effect of changes in the model inputs \cite{Borgonovo2016}. More explicitly, sensitivity analysis determines the key drivers for uncertainty in the model output.
In the following $\boldsymbol{Y}(\theta) \in \Xi \subset \mathbb{R}^M$ denotes the vector of $M$ random input variables and $\mathcal{Q}(\boldsymbol{Y}(\theta)) \in \Xi_{\mathcal{Q}} \subset \mathbb{R}$ represents the model output, named \textit{quantity of interest} (QoI) hereafter. Note that in this work the inputs $\boldsymbol{Y}$ refer to the geometrical parameters specified in \eqref{eq:paravec} while the output $\mathcal Q$ is given by the cell-to-cell coupling coefficient $k_\text{cc}$.

Computing derivatives of the QoI is a frequently used approach for sensitivity analysis, however, especially in the case of non-linear models this approach only provides locally information at the nominal value where the derivatives are computed. Therefore, global sensitivity analysis (GSA) methods are developed, that explore the effect of changes in the whole input space \cite{Saltelli2008}.

Standard indices used for GSA are Sobol indices \cite{saltelli2000sensitivity}, which require independent input parameters. Contrary to Sobol indices, Borgonovo indices \cite{Borgonovo_2007} do not rely on this assumption. Moreover, they are moment independent quantities. Both indices are explained in more detail in the next two subsections.  

\subsection{Sobol indices} 
 Sobol indices belong to the class of variance-based sensitivity analysis and are based on the idea of decomposing the variance of the QoI. A detailed description and derivation of the Sobol indices is given in \cite{Saltelli2008} and a brief summary, following the exposition in \cite{Kadyk2019}, is given hereafter. In a first step, the QoI is decomposed by a so called Hoeffding decomposition as
\begin{align}
\mathcal{Q}(\boldsymbol{Y}) &= \mathcal{Q}_0 + \sum_{i=1}^M \mathcal{Q}_i(Y_i) + \sum_{i=1}^M \sum_{j > i} \mathcal{Q}_{i,j}(Y_i,Y_j) + \dots \\ \nonumber &+ \mathcal{Q}_{1,2,\dots,M}(Y_1, \dots, Y_M),
\end{align}
where 
\begin{align}
\mathcal{Q}_0 &= \mathbb{E} [ \mathcal{Q} ], \\
\mathcal{Q}_i(Y_i) &= \mathbb{E}_{Y \sim\, i} [ \mathcal{Q} \vert Y_i] - \mathbb{E} [Q], \\
\mathcal{Q}_{i,j}(Y_i,Y_j) &= \mathbb{E}_{Y \sim\, i,j} [ \mathcal{Q} \vert Y_i, Y_j] - \mathcal{Q}_i - \mathcal{Q}_j - \mathbb{E} [Q]. 
\end{align}
Here, $\mathbb{E}_{Y \sim\, i} [ \mathcal{Q} \vert Y_i]$ denotes the expected value, conditional on the random variable $Y_i$. In the presence of independent input random variables this orthogonal function decomposition can be transformed into a decomposition of the variance as
\begin{align}
\label{eq:ANOVA}
\mathbb{V}[ \mathcal{Q}] &= \sum_{i=1}^{M} \mathbb{V}[\mathcal{Q}_i (Y_i)] + \sum_{i = 1}^{M} \sum_{j > 1} \mathbb{V} [\mathcal{Q}_{i,j}(Y_i,Y_j)] + \dots \\ \nonumber &+ \mathbb{V}[\mathcal{Q}_{1,2,\dots,M}(Y_1,\dots,Y_M)].
\end{align} 

Equation (\ref{eq:ANOVA}) can be interpreted as follows: The first sum in (\ref{eq:ANOVA}) describes the individual contribution of the input parameter $Y_i$ to the overall variance $\mathbb{V}[\mathcal{Q}]$, while the other terms of the sum contain contributions from combined parameter effects.

Based on this decomposition the first order Sobol indices are defined as 
\begin{equation}
S_i = \frac{\mathbb{V}[\mathcal{Q}_i(Y_i)]}{\mathbb{V}[\mathcal{Q}]}.
\end{equation}

In addition to the first order indices there are total effect indices $S^T_i$, which describe the contribution of an individual parameter including higher order (interaction) effects with other parameters to the total variance. As a consequence, by comparing the total indices with the first order indices the modeller can detect potential interaction effects among the input parameters, i.e., if $S^T_i > S_i$. \\

One approach to compute the Sobol indices is a Monte Carlo based sampling approach introduced in \cite{Saltelli2008} and often referred to as Saltelli's algorithm. Basically, two independent samples of the input vector $\boldsymbol{Y}$ are generated. Each sample contains $N$ realizations of each input variable and both samples are stored in a $M \times N$ matrix $\mathbf{A}$ and $\mathbf{B}$, respectively. Additionally a matrix $\mathbf{C}_i$ is created, by keeping the i-th row of matrix $\mathbf{A}$ and replacing all other rows by the entries of matrix $\mathbf{B}$. The matrix $\mathbf{C}_i$ will be used to approximate the conditional expectation of the QoI, by fixing the variation in one parameter.
The first order Sobol index can then be approximated by
\begin{equation}
S_i \approx \frac{\frac{1}{N}\sum_{j=1}^N \mathcal{Q}(\mathbf{a}_j) \mathcal{Q}(\mathbf{c}_j) - \mathbb{E}^{\text{MC}}_{\mathbf{A}}[\mathcal{Q}]^2}{\mathbb{V}^{\text{MC}}_{\mathbf{A}}[\mathcal{Q}]},
\end{equation}
where $\mathbb{E}^{\text{MC}}_{\mathbf{A}},\mathbb{V}^{\text{MC}}_{\mathbf{A}}$ denote the Monte Carlo approximation of the mean value and variance based on the sample stored in $\mathbf{A}$, respectively. Also, $\mathbf{a}_j$ and $\mathbf{c}_j$ denote one column of the respective sample matrix, which is equal to one independently drawn realization. For further details and the approximation of the total order indices, the reader is referred to \cite{Saltelli2008}.

\subsection{Borgonovo indices}
The Sobol indices presented in the last section are global and quantitative indicators, which do not assume linearity of the model. However, Sobol indices belong to the class of variance based methods that "implicitly assume that this moment (variance) is sufficient to describe the output variability" \cite{saltelli2002sensitivity}. In addition, Sobol indices require, that the input variables are statistically independent. 
 
To overcome these limitations Borgonovo \cite{Borgonovo_2007} introduced a new moment-independent importance measure, which analyses the effect of each input on the entire output distribution. To this end, the difference between the unconditional pdf $\rho_{\mathcal{Q}}$ and the conditional pdf $\rho_{{\mathcal{Q}}\vert Y_i}$ is computed by the shift function 
\begin{equation}
s(Y_i) = \int_{\Xi_{\mathcal{Q}}} \vert \rho_{\mathcal{Q}}(q) - \rho_{{\mathcal{Q}}\vert Y_i}(q)\vert \,\mathrm{d}q.
\end{equation}
The unconditional pdf $\rho_{\mathcal{Q}}$ represents the output distribution, when all input parameter are allowed to vary and the conditional pdf $\rho_{{\mathcal{Q}}\vert Y_i}$ represents the case when the uncertainty in one input is eliminated by fixing it to one value. 

Taking the expectation of the shift function w.r.t. $Y_i$ and multiplying by a normalization constant yields the Borgonovo sensitivity indices
\begin{align}
\label{eq:Borgonovo}
\delta_i &= \frac{1}{2} \mathbb{E}_{Y_i}[s(Y_i)] \nonumber \\
 &= \frac{1}{2} \int_{\Xi} \rho_{Y_i} (y_i) \bigg[ \int_{\Xi_{\mathcal{Q}}} \vert \rho_{\mathcal{Q}}(q) - \rho_{{\mathcal{Q}}\vert Y_i = y_i}(q)\vert \,\mathrm{d}q \bigg] \,\mathrm{d}y_i.
\end{align}
In fact it is shown in \cite{da2015global} that the Borgonovo sensitivity indices are a special case of a global sensitivity analysis framework based on a $f$-divergence dissimilarity measure between the unconditional and conditional distribution of ${\mathcal{Q}}$.

In \cite{Borgonovo_2007} several properties of the Borgonovo indices are proven. First of all, the indices are normalized and bounded such that $0 \leq \delta_i \leq 1$. If $\mathcal{Q}$ is independent on $Y_i$ then $\delta_i = 0$. Since no independence of the input variables is required, the Borgonovo indices can also be applied in the presence of dependent inputs.

Following \cite{da2015global}, the Borgonovo indices can be computed by reformulating (\ref{eq:Borgonovo}) in terms of marginal and joint probability densities as
\begin{equation}
\label{eq:BorgonovoDensities}
\delta_i = \frac{1}{2}\int_{\Xi \times \Xi_{\mathcal{Q}}} \vert \rho_{\mathcal{Q}}(q) \rho_{Y_{i}}(y_i) - \rho_{{\mathcal{Q}},Y_i = y_i}(q,y_i) \vert \,\mathrm{d}q \,\mathrm{d}y_i. 
\end{equation}

In this work the joint and marginal densities are estimated via kernel-density estimators based on the given data sample with automated band width selection. The integral in (\ref{eq:BorgonovoDensities}) is then estimated via a Monte Carlo approach, where $y^{(j)}_i,q^{(j)}$ denote a sample point of the corresponding input/output distribution:  
\begin{equation}
\label{eq:BorgonovoMC}
\delta_i \approx \frac{1}{2}\frac{1}{n}\sum_{j = 1}^{n} \vert \rho_{\mathcal{Q}}(q^{(j)}) \rho_{Y_{i}}(y^{(j)}_i) - \rho_{{\mathcal{Q}},Y_i}(q^{(j)},y^{(j)}_i) \vert.
\end{equation}

The indices computed by (\ref{eq:BorgonovoMC}) are verified via a second approach, which approximates (\ref{eq:Borgonovo}) by a histogram-based approach and a good agreement of the two approaches is observed. See \cite{Borgonovo_2007} for further information on the histogram-based approach. 
\section{Surrogate Modeling}
\label{sec:surrogate}

As discussed in Section~\ref{sec:uncertainty}, the proposed uncertainty quantification study requires the repeated solution of the parametrized eigenvalue problem \eqref{eq:Maxwell-eig-cont} for different inputs $\boldsymbol{Y}$. In order to achieve a high accuracy, a large number of random (\textit{virtual}) cavities $\tilde N_\text{cav}$ is desirable. However, in this case, the computational cost of repeatedly solving \eqref{eq:Maxwell-eig-cont} with the finite element method would become prohibitive. Hence, we employ surrogate modeling to keep the computational effort feasible. The main idea of polynomial surrogate modeling is to approximate the mapping from input parameters $\boldsymbol{y}\in \Xi\subset\mathbb R^M$, which might, for example, represent a realization of the random vector given in \eqref{eq:paravec}, to an output quantity $\mathcal Q$, which, e.g., might refer to $f_1,\,\ldots,f_9$ (fundamental mode spectrum) or  $k_\text{cc}$, using global polynomial approximations. In this case spectral convergence can be expected \cite{xiu2010}. In particular, we use approximations of the form 
\begin{equation}
    \mathcal Q(\boldsymbol{y})
    \approx \sum_{i=0}^N 
    c_i
    \Psi_i(\boldsymbol{y})=:\mathcal Q_N(\boldsymbol{y}),  \label{eq:poly_approx}
\end{equation}
where $\Psi_i$ 
are global multivariate polynomials and $c_i$ denote the associated coefficients. In this work, we rely on the stochastic collocation method \cite{nobile2008sparse} to compute the approximation \eqref{eq:poly_approx}. In particular, we use the algorithm proposed in \cite{narayan2014}, with minor modifications to address the case of multiple quantities of interest. In the following, we recall the main ideas but refer to \cite{narayan2014, loukrezis2019accessing} for the specific details. 

The method is based on Leja nodes \cite{Leja1957} which are nested interpolation nodes defined by an optimization problem, s.t. for example the uni-variate nodes $\{y_1^{(i)}\}_{i}$ are given recursively by 
\begin{equation}
    y_1^{(I)} = \arg \min_{y_1} \prod_{i=0}^{I-1} |y_1 - y_1^{(i)}|. \label{eq:leja_nodes}
\end{equation}
Since Leja nodes allow, by construction, for a nested and granular refinement, they are well suited for adaptive approximations in the multivariate case \cite{narayan2014, loukrezis2019adaptive}.
Note that there are also generalizations of \eqref{eq:leja_nodes}, i.e., \textit{weighted} Leja nodes tailored to the probability density of the input \cite{narayan2014, loukrezis2019approximation}. However, those are not considered here, as we aim for high uniform accuracy. 

Corresponding multivariate nodes $\{\boldsymbol{y}^{(i)}\}_i$ could be obtained by a tensor-product construction of univariate nodes, i.e. 
\begin{equation}
    \{y_1^{(i)}\}_{i}\times \{y_2^{(i)}\}_{i}\times\ldots \times \{y_M^{(i)}\}_{i}. \label{eq:tensor_grid}
\end{equation}
Stochastic collocation then consists in evaluating the FE model for all $\boldsymbol{y}^{(i)}$ and enforcing the corresponding collocation conditions for the surrogate model, i.e.
\begin{equation}
    \mathcal Q(\boldsymbol{y}^{(i)})
    \overset ! =   \mathcal Q_N(\boldsymbol{y}^{(i)}) \quad \forall \boldsymbol{y}^{(i)}.
\end{equation}
For  a larger number of parameters (e.g. $M>4$) the computational cost associated to the tensor grid \eqref{eq:tensor_grid} would become prohibitive. Hence, the employed algorithm constructs a sparse-grid, cf. \cite{bungartz2004sparse}, which utilizes only a subset of \eqref{eq:tensor_grid} and neglects points which do not significantly contribute to the accuracy of the approximation. To this end, the adaptive selection of admissible nodes for refinement is based on the respective point-wise error 
$\epsilon = |\mathcal Q-\mathcal Q_N|$ if a scalar output quantity is considered. The node associated to the largest error is then chosen to refine the approximation. If several quantities  of interest $\mathcal Q^{(1)}, \mathcal Q^{(2)}, \ldots$ shall be approximated simultaneously based on the same set of nodes, we replace $\epsilon$ by \begin{equation}
\tilde \epsilon = \max \{w_1|\mathcal Q^{(1)}-\mathcal Q^{(1)}_N|, w_2|\mathcal Q^{(2)}-\mathcal Q^{(2)}_N|, \ldots\},
\end{equation}
where $w = [w_1,w_2,\ldots]$ denotes a vector of weights.
The adaptive selection of interpolation nodes based on $\tilde \epsilon$ is terminated when a given computational budget is reached. For details on the definition of admissible nodes based on downward-closed multi-index sets and the employed multivariate hierarchical Lagrange polynomials, we refer to \cite{narayan2014, loukrezis2019accessing}. 

First, the surrogate modeling technique is applied to construct the 4-variate polynomial surrogate models for step 3 of Section~\ref{sec:uncertainty}. In particular, using a computational budget of 50 FE model evaluations, we construct approximations of the fundamental eigenfrequency $f$ of cell $i$ w.r.t. the variations in both adjacent iris radii $\Delta R_\text{ir}^{(i)}, \Delta R_\text{ir}^{(i+1)}$, the equator radii $\Delta R_\text{eq}^{(i)}$ and the length $\Delta L^{(i)}$.
The accuracy of the approximations can then be quantified by cross-validation in terms of the empirical $L^\infty$ norm. In particular, for a set of random parameter realizations $\{\boldsymbol{y}^{\text{cv}}\}_{i=1}^{N^\text{cv}}$, we compute
\begin{equation}
E^\text{cv} = \max_{\boldsymbol{y} \in \{\boldsymbol{y}^{\text{cv}}\}_{i=1}^{N^\text{cv}}} \bigl|\mathcal Q(\boldsymbol{y})-\mathcal Q_N(\boldsymbol{y})\bigr|.
\end{equation}
For the end-cells as well as the middle-cells, the error $E^\text{cv}$, computed with a uniformly distributed sample of size $N^\text{cv}=1000$, is below \SI{5}{kHz}.

Next, we compute the 19-variate surrogate models employed in step 8 using 500 FE model evaluations. In this case, we consider the 19-dimensional input vector \eqref{eq:paravec} and 10 output quantities $f_1,\ldots, f_9, k_\text{cc}$ with weight vector $w=[1,\ldots,1,0]$. Cross-validation for the sample $\{\boldsymbol{Y}_\text{sort}^{(m)}\}_{m=1}^{1000}$, see Section~\ref{sec:uncertainty}, indicates an error below \SI{10}{kHz} for all fundamental eigenfrequencies and of less than $5\cdot 10^{-4}$ for the cell-to-cell coupling coefficient $k_\text{cc}$.

The estimates for expectation as well as standard deviation for each output quantity $\mathcal Q^{(i)}$ are then obtained from the respective surrogate models $\mathcal Q_N^{(i)}$ as
\begin{align}
    \mathbb E[Q^{(i)}] &\approx \frac 1 {\tilde N_\text{sel}} \sum_{m=1}^M \mathcal Q_N^{(i)}\bigl(\boldsymbol{Y}_\text{sort}^{(m)}\bigr) := \mathbb E^\text{MC}[Q^{(i)}],\\
    \operatorname{Std}[Q^{(i)}]^2 &\approx \frac  {\sum_{m=1}^{\tilde N_\text{sel}} \left(\mathcal Q_N^{(i)}\bigl(\boldsymbol{Y}_\text{sort}^{(m)}\bigr)- \mathbb E^\text{MC}[Q^{(i)}]\right)^2} {\tilde N_\text{sel}-1}.
\end{align}

}

\end{document}